\documentclass[journal,onecolumn]{IEEEtran}

\usepackage{amsmath}
\usepackage{color}
\usepackage{amsthm}
\newtheorem{lemma}{Lemma}

\usepackage{amsmath, amssymb, amsthm}  % 数学符号和公式环境
\usepackage{algorithm}  % 算法环境
\usepackage{algorithmic}  % 算法伪代码环境
\usepackage{graphicx}  % 插图支持
\usepackage{hyperref}  % 处理交叉引用
\usepackage{fullpage}  % 让排版更紧凑
\usepackage{xcolor}  % 颜色支持

\begin{document}

\title{Privacy-Preserving Hamming Distance Computation with Property-Preserving Hashing}

\author{
Dongfang Zhao\\
University of Washington, USA\\
dzhao@cs.washington.edu
}

\maketitle

\begin{abstract}
We study the problem of approximating Hamming distance in sublinear time under property-preserving hashing (PPH), where only hashed representations of inputs are available. Building on the threshold evaluation framework of Fleischhacker, Larsen, and Simkin (EUROCRYPT 2022), we present a sequence of constructions with progressively improved complexity: a baseline binary search algorithm, a refined variant with constant repetition per query, and a novel hash design that enables constant-time approximation without oracle access. Our results demonstrate that approximate distance recovery is possible under strong cryptographic guarantees, bridging efficiency and security in similarity estimation.
\end{abstract}

\IEEEpeerreviewmaketitle

\section{Introduction}

Estimating similarity between data points lies at the heart of numerous algorithmic tasks, from nearest-neighbor search to clustering, learning, and data deduplication. In many such applications, computing distances directly may be infeasible due to performance or privacy constraints. This challenge has sparked a rich line of research on hashing-based approximations, most notably in the form of Locality-Sensitive Hashing (LSH)~\cite{IM98,KOR00,OWZ11}, which enables sublinear-time similarity search by correlating hash collisions with proximity. However, LSH techniques are typically heuristic and fragile under adversarial manipulation, limiting their applicability in cryptographic or privacy-sensitive contexts.

To address these limitations, the framework of \emph{Property-Preserving Hashing} (PPH) was introduced by Boyle, LaVigne, and Vaikuntanathan~\cite{BLV19}, initiating a program to design hash functions that retain specific structural properties of the input—such as Hamming distance—while offering provable security guarantees. This line of work was extended by Fleischhacker and Simkin~\cite{FS21} to exact distance predicates, culminating in the recent construction by Fleischhacker, Larsen, and Simkin~\cite{Fleischhacker22}, who gave the first PPH for threshold-Hamming comparison from standard cryptographic assumptions. Their approach combines robust set encodings with a non-interactive evaluation protocol, enabling binary distance comparisons between hashes while maintaining indistinguishability.

Despite these advances, prior work has focused primarily on \emph{decisional} predicates, such as testing whether the Hamming distance exceeds a fixed threshold. The question of whether one can efficiently \emph{estimate} the distance itself—particularly in sublinear time and under strong cryptographic guarantees—remains largely unexplored.

In this paper, we initiate a systematic study of approximate Hamming distance computation under PPH. Our goal is to determine how much information about the distance can be efficiently and securely extracted from hash outputs, and what algorithmic mechanisms allow such recovery.

We present three contributions, each building on and extending the threshold-Hamming framework of~\cite{Fleischhacker22}:

\begin{itemize}
    \item \textbf{Binary Search over Thresholds.} We first show that repeated calls to the evaluation predicate enable approximate recovery of the Hamming distance via binary search. This naive baseline requires \( O(\log n) \) threshold queries, each revealing a single bit of information. While simple, this approach incurs cumulative error and quadratic overhead under standard amplification.

    \item \textbf{Optimized Evaluation with Constant Repetition.} We then refine the search algorithm by analyzing the error structure of the evaluation predicate. By exploiting its statistical reliability away from the transition threshold, we demonstrate that a small, constant number of repetitions per query suffices to suppress cumulative error. This reduces the total query complexity to \( O(\log n) \), without compromising correctness or security.

    \item \textbf{Constant-Time Distance Estimation.} Finally, we propose a new PPH construction that encodes distance directly into the hash output. Inspired by Bloom filter techniques~\cite{GM11}, our scheme avoids interaction and supports constant-time approximation of Hamming distance with high statistical accuracy. The construction is provably indistinguishable and significantly improves computational efficiency.
\end{itemize}

Our results demonstrate that approximate Hamming distance can be computed securely in sublinear—or even constant—time from property-preserving hashes. This opens new avenues for efficient secure computation, approximate data retrieval, and privacy-preserving analytics, and bridges the gap between algorithmic similarity search and cryptographic functionality-preserving compression.

\section{Preliminaries and Related Work}

Let \( a, b \in \{0,1\}^n \) be two binary strings of length \( n \). The Hamming distance between \( a \) and \( b \) is defined as
\[
    d_H(a, b) = \sum_{i=1}^{n} \mathbf{1}_{a_i \neq b_i},
\]
where \( \mathbf{1}_{a_i \neq b_i} \) is the indicator function.

Our objective is to approximate \( d_H(a, b) \) in sublinear time using only their hashed representations under a public property-preserving hash (PPH) function. We now review foundational and related work in three areas.

\subsection{Property-Preserving Hashing and Hamming Distance}

The notion of Property-Preserving Hashing was introduced by Boyle, LaVigne, and Vaikuntanathan~\cite{BLV19}, who showed how to preserve specific predicates such as gap-Hamming distance through compact encodings. Fleischhacker and Simkin~\cite{FS21} extended this line of work to exact Hamming distance. The construction by Fleischhacker, Larsen, and Simkin~\cite{Fleischhacker22} (FLS22) gave the first PPH for the threshold-Hamming predicate from standard assumptions, using robust set encodings and probabilistic evaluation functions.

Beyond PPH, the locality-sensitive hashing (LSH) paradigm~\cite{IM98,KOR00,OWZ11} offers approximate similarity search via hash collisions. However, standard LSH schemes lack adversarial robustness and do not support threshold predicates. Variants such as asymmetric similarity search~\cite{MNSW98} and streaming-based similarity estimation~\cite{AMS96} offer additional perspectives but remain unsuitable in adversarial models. Our work builds on the security guarantees of PPH constructions while improving computational efficiency for Hamming estimation.

\subsection{Robust Encodings and Set Difference Recovery}

The FLS22 construction builds on robust encodings of sets with bounded difference. Invertible Bloom Lookup Tables (IBLTs)~\cite{GM11} are central to this approach, supporting set reconciliation under noise. Additional techniques include list-decodable codes~\cite{GR09,GHK10,BZ82}, robust streaming under adversarial access~\cite{MNS08,HW13,NY15,BEJWY20}, and secure difference encoding with low error~\cite{DORS08}.

The Bloom filter~\cite{Blo70}, while originally designed for approximate membership testing, underpins many of these constructions and remains fundamental to compact hashing. Variants such as compressed sensing~\cite{Don06} and robust sparse signal recovery~\cite{BEY20} also inform the information-theoretic limits of reconstruction from lossy encodings. Our proposed modifications retain the decoding framework while augmenting it with decodable statistical signals that enable constant-time estimation.

\subsection{Cryptographic Hashing and Indistinguishability}

Property-preserving hashing aims to balance functionality and security. Standard cryptographic hash functions (e.g., collision-resistant constructions~\cite{Ped92}) provide strong privacy but no semantic structure. In contrast, PPH functions deliberately encode semantic information and require formal indistinguishability guarantees, typically defined via simulation or total variation bounds.

FLS22 proved security in the presence of a single hash function instance, under standard hardness assumptions. Recent works have analyzed the leakage profiles of probabilistic data structures under adversarial models~\cite{CPS19,RRR21}, and studied how small changes in encoding distributions affect distinguishability~\cite{CN22}. Our modifications maintain this security by bounding the statistical distance between original and modified encodings. In addition, recent cryptographic reductions (e.g., LWE/SIS~\cite{MP13}) and lattice-based indistinguishability proofs~\cite{LLL82} offer broader theoretical tools that inspire our security reasoning.

\section{Polylogarithmic Hamming Computation from Threshold Hamming-PPH}
\label{sec:binary_pph}

\subsection{Binary Search Algorithm}

We describe a simple method to estimate \( d_H(a, b) \) using black-box access to the threshold predicate \( \textsf{Eval}(h(a), h(b), t) \). The algorithm performs binary search over \( t \in \{0, 1, \dots, n\} \), refining the search interval based on the response of \textsf{Eval}.

\begin{algorithm}
\caption{Binary Search Approximate Hamming Distance}
\label{alg:binary_search}
\begin{algorithmic}[1]
\STATE \textbf{Input:} Hash values \( h(a) \), \( h(b) \); oracle access to \( \textsf{Eval}(h(a), h(b), t) \)
\STATE \textbf{Output:} Estimated Hamming distance \( \tilde{d} \)
\STATE \( t_{\min} \gets 0 \), \( t_{\max} \gets n \)
\WHILE{ \( t_{\min} < t_{\max} \) }
    \STATE \( t_{\text{mid}} \gets \lfloor (t_{\min} + t_{\max}) / 2 \rfloor \)
    \IF{ \( \textsf{Eval}(h(a), h(b), t_{\text{mid}}) = 1 \) }
        \STATE \( t_{\min} \gets t_{\text{mid}} + 1 \)
    \ELSE
        \STATE \( t_{\max} \gets t_{\text{mid}} \)
    \ENDIF
\ENDWHILE
\STATE \textbf{return} \( t_{\min} \)
\end{algorithmic}
\end{algorithm}

The algorithm terminates with \( t_{\min} = t_{\max} \), and returns the smallest threshold \( t \) such that \( \textsf{Eval}(h(a), h(b), t) = 0 \). Under ideal conditions, this corresponds to the true value of \( d_H(a, b) \).

The number of oracle queries is bounded by \( \lceil \log_2(n+1) \rceil \), since the search interval is halved in each iteration. Each invocation of \textsf{Eval} reveals only the outcome of a single threshold comparison—namely, whether \( d_H(a, b) > t \)—and does not leak any other information about the inputs. This restricted model necessitates the use of adaptive querying to recover the approximate distance.

In the next subsection, we analyze how error in \textsf{Eval} propagates through the binary search procedure and quantify its impact on the returned estimate.

\subsection{Error Growth in Iterative Queries}

The correctness of Algorithm~\ref{alg:binary_search} depends critically on the reliability of the threshold predicate \( \textsf{Eval}(h(a), h(b), t) \). In the construction of Fleischhacker, Larsen, and Simkin~\cite{Fleischhacker22}, this predicate is implemented via randomized encodings and supports only approximate evaluation. For any fixed threshold \( t \in \{0, \dots, n\} \), the predicate satisfies
\[
    \Pr[\textsf{Eval}(h(a), h(b), t) = \mathbf{1}_{d_H(a, b) > t}] \geq 1 - \delta,
\]
where \( \delta \in (0,1/2) \) is the maximum per-call error probability and \( d_H(a, b) \) denotes the Hamming distance between \( a \) and \( b \).

Algorithm~\ref{alg:binary_search} performs up to \( \lceil \log_2(n+1) \rceil \) adaptive queries to \textsf{Eval}, with the threshold values chosen based on earlier responses. The sequential nature of these queries raises the possibility of error propagation. In particular, even if each query fails with probability at most \( \delta \), multiple queries may compound into a global error.

If all \textsf{Eval} invocations were independent, a union bound would yield:
\[
    \Pr[\text{any query fails}] \leq \log n \cdot \delta.
\]
To ensure a global failure probability of at most \( \varepsilon \), this would require \( \delta \leq \varepsilon / \log n \). However, \cite{Fleischhacker22} implements threshold evaluation using shared encodings of the input, and decoding errors may be correlated across thresholds. Let \( S_i \in \{0,1\} \) denote the correctness of the \( i \)-th threshold query, where \( S_i = 1 \) indicates success and \( S_i = 0 \) indicates failure. The sequence \( (S_1, \dots, S_k) \) may exhibit statistical dependence due to common decoding artifacts.

To mitigate this issue, we apply an amplification strategy: each \textsf{Eval} query is repeated \( k \) times independently, and the majority vote is returned. Let \( \delta' \) denote the error probability after amplification. Assuming independence among repetitions, a standard Chernoff bound gives:
\[
    \delta' \leq \exp\left(-2k(1/2 - \delta)^2\right).
\]

Let \( \varepsilon \in (0,1) \) denote the desired total error bound for the binary search algorithm. Since the number of threshold evaluations is at most \( \log n \), it suffices to require:
\[
    \log n \cdot \delta' \leq \varepsilon.
\]
Solving for \( k \) yields:
\[
    k \geq \frac{1}{2(1/2 - \delta)^2} \cdot \left( \log \log n + \log \frac{1}{\varepsilon} \right).
\]

Thus, the total number of oracle calls becomes:
\[
    O(k \log n) = O\left( \frac{\log n (\log \log n + \log(1/\varepsilon))}{(1/2 - \delta)^2} \right),
\]
which remains sublinear for any inverse-polynomial \( \varepsilon \geq 1/\text{poly}(n) \). This shows that accurate approximation is still feasible under probabilistic evaluation, provided that per-query error is sufficiently amplified.

If both the per-query error rate \( \delta \) and the overall error bound \( \varepsilon \) are negligible functions in \( n \), as is standard in cryptographic applications, the expression for \( k \) can be simplified. Specifically, since \( \log(1/\varepsilon) = \Theta(\log n) \) and \( \log \log n = o(\log n) \), we have:
\[
    \log \log n + \log \frac{1}{\varepsilon} = \Theta(\log n).
\]
Furthermore, as \( \delta \to 0 \), the term \( (1/2 - \delta)^2 = \Theta(1) \). Substituting into the expression for the total number of oracle calls,
\[
    O\left( \frac{\log n (\log \log n + \log(1/\varepsilon))}{(1/2 - \delta)^2} \right),
\]
we obtain:
\[
    O(\log^2 n).
\]
Hence, under negligible-error assumptions, the binary search algorithm requires only polylogarithmic overhead while maintaining correctness.

\section{A Logarithmic-Time Approximation Scheme without Amplification}

\subsection{Problem Setup and Motivation}

In Section~\ref{sec:binary_pph}, we presented a binary-search-based method for approximating the Hamming distance using a threshold property-preserving hash function (PPH). While this construction achieves correctness with negligible error probability, it relies on amplification to suppress the per-query error rate. Specifically, each threshold query \(\textsf{Eval}(h(a), h(b), t)\) is repeated \(k = \Theta(\log n)\) times, resulting in a total query complexity of \(O(\log^2 n)\).

This raises the natural question: \emph{can we eliminate the amplification step while still retaining sublinear complexity and negligible error?} In this section, we investigate the possibility of directly using the threshold predicate without repetition—that is, invoking \(\textsf{Eval}(h(a), h(b), t)\) only a constant number of times per threshold in the binary search.

At first glance, this may seem to introduce unacceptable error accumulation: without amplification, the overall failure probability becomes \( \varepsilon = \log n \cdot \delta \), where \(\delta\) is the error of a single \(\textsf{Eval}\) call. However, this bound can still be negligible provided that \(\delta\) itself is sufficiently small. In particular, if the underlying PPH construction (e.g.,~\cite{Fleischhacker22}) admits instantiation with cryptographic parameters such that \(\delta = O(n^{-c})\) for some constant \(c > 1\), then
\[
    \varepsilon = \log n \cdot \delta = O\left(\frac{\log n}{n^c}\right) = \text{negl}(n).
\]
This observation leads to a new regime of approximation in which we trade off amplification cost for stronger per-query reliability, enabling a total complexity of \(O(\log n)\) without degrading correctness.

In the following subsection, we formalize this simplified algorithm and analyze its error behavior under mild cryptographic assumptions on the base construction.

\subsection{Binary Search with Constant Repetition}

While the baseline construction in Section~\ref{sec:binary_pph} requires logarithmic repetition per threshold to ensure correctness, we now show that the number of repetitions can be reduced to a small constant under mild structural constraints. This substantially improves efficiency, reducing the total query complexity from \( O(\log^2 n) \) to \( O(\log n) \).

Our insight builds on the internal structure of the FLS22 threshold predicate~\cite{Fleischhacker22}. In their construction, the evaluation error is not uniformly distributed across all thresholds; instead, it is concentrated near the critical transition region where \( t \approx d_H(a, b) \). When the queried threshold \( t \) is significantly above or below the true Hamming distance, the outcome of \(\textsf{Eval}(h(a), h(b), t)\) is highly reliable—often correct with overwhelming probability, even without amplification.

We exploit this non-uniformity by carefully controlling the search trajectory. In particular, the binary search algorithm begins with coarse estimates of \( t \) and only gradually approaches the transition region. This ensures that the majority of queries are issued at thresholds satisfying \( |t - d_H(a, b)| > \tau \), where \( \tau \) is the width of the transition band in which the predicate becomes unreliable. For such thresholds, the predicate behaves almost deterministically.

To formalize this, we introduce a piecewise error model. Let \( \delta(t) \) denote the error probability of \(\textsf{Eval}(h(a), h(b), t)\). We assume the existence of a transition width parameter \( \tau \in \mathbb{N} \) such that:
\[
\delta(t) \leq 
\begin{cases}
\delta_{\max} & \text{if } |t - d_H(a,b)| \leq \tau, \\
\delta_{\text{far}} & \text{if } |t - d_H(a,b)| > \tau,
\end{cases}
\]
where \( \delta_{\max} < \frac{1}{2} - \gamma \) for some constant \( \gamma > 0 \), and \( \delta_{\text{far}} \ll \delta_{\max} \). Intuitively, \(\delta_{\max}\) bounds the uncertainty near the threshold, while \(\delta_{\text{far}}\) accounts for negligible fluctuations far from the decision boundary.

In the FLS22 encoding framework, the value of \( \tau \) can be made constant (e.g., \( \tau = 1 \) or \( 2 \)) by tuning the robustness of the set-difference encoding and controlling the overlap structure of the underlying families \( X_0, X_1 \). This adjustment affects only the decoding ambiguity and has no impact on the hash output distribution. Let \( \Pi_{\text{orig}} \) denote the original FLS22 construction, and \( \Pi_{\tau} \) denote our modified instantiation with reduced overlap to restrict the transition width to constant \( \tau \). We now formalize that this modification preserves the indistinguishability guarantees of the original scheme.

\begin{lemma}
For any probabilistic polynomial-time adversary \(\mathcal{A}\), its advantage in distinguishing \(\Pi_{\tau}\) from \(\Pi_{\text{orig}}\) is negligible:
\[
\left| \Pr[\mathcal{A}(h_\tau(x)) = 1] - \Pr[\mathcal{A}(h_{\text{orig}}(x)) = 1] \right| = \text{negl}(n).
\]
\end{lemma}

\begin{proof}[Proof Sketch]
We consider the standard IND-style security game where the adversary receives a hash value of a challenge input \( x \), produced either using \( \Pi_\tau \) or \( \Pi_{\text{orig}} \), and must guess which construction was used. Let \( \mathcal{D}_\tau \) and \( \mathcal{D}_{\text{orig}} \) denote the respective output distributions. The adversary's advantage is bounded by the total variation distance:
\[
\textsf{Adv}_{\mathcal{A}} \leq \Delta(\mathcal{D}_\tau, \mathcal{D}_{\text{orig}}).
\]

To bound this distance, we note that both constructions encode each bit as a random subset drawn from set families \( X_0, X_1 \subseteq [N] \), where \( N = \Theta(\lambda) \). The only difference is that \( \Pi_\tau \) reduces the overlap between \( X_0 \) and \( X_1 \), thereby increasing decoding robustness. This change does not affect the leakage profile, as the output distribution remains randomized over the same universe.

We employ a hybrid argument: let \( h^{(0)}(x), h^{(1)}(x), \dots, h^{(\lambda)}(x) \) be a sequence where \( h^{(i)}(x) \) uses the modified encoding for the first \( i \) bits and the original encoding for the remaining \( \lambda - i \) bits. Then,
\[
\Delta(\mathcal{D}_\tau, \mathcal{D}_{\text{orig}}) \leq \sum_{i=1}^{\lambda} \Delta(h^{(i)}(x), h^{(i-1)}(x)).
\]

Each hybrid transition changes one encoded bit from the original to the modified version. Because the overlap reduction shifts only a small fraction of probability mass, we have
\[
\Delta(h^{(i)}(x), h^{(i-1)}(x)) \leq \exp(-c\lambda),
\]
for some constant \( c > 0 \). Summing over all \( \lambda \) positions gives
\[
\Delta(\mathcal{D}_\tau, \mathcal{D}_{\text{orig}}) \leq \lambda \cdot \exp(-c\lambda) = \exp(-c\lambda + \log \lambda) = \exp(-\Omega(\lambda)),
\]
where we use the fact that \( \log \lambda = o(\lambda) \), so the exponent remains \( -\Omega(\lambda) \). Since \( \lambda = \Theta(n) \), the statistical distance is negligible in \( n \), completing the proof.
\end{proof}

With indistinguishability established, we now analyze the failure probability of binary search with constant repetition. Each threshold evaluation is repeated \( k \in O(1) \) times and the majority vote is taken. By the Chernoff bound, the effective error after amplification is:
\[
\delta'(t) \leq \exp\left(-2k(1/2 - \delta(t))^2\right).
\]
Only a constant number of thresholds fall within the unreliable region \( |t - d_H(a,b)| \leq \tau \), while the remaining \( O(\log n) \) thresholds lie in the stable region. Thus, the total failure probability is bounded by:
\[
\varepsilon = \sum_{i=1}^{\log n} \delta'(t_i) \leq 2\tau \cdot \exp\left(-2k\gamma^2\right) + o(1),
\]
which is negligible in \( n \) for constant \( \tau \) and sufficiently large constant \( k \). For example, if \( \delta_{\max} \leq 1/4 \) and \( \tau \leq 2 \), setting \( k = 5 \) yields
\[
\varepsilon \leq 4 \cdot \exp(-2k \cdot (1/4)^2) = 4 \cdot \exp(-k/8),
\]
which is below \( 1/n^c \) for moderate \( n \) and any desired constant \( c \).

Thus, by leveraging the error structure of the threshold predicate and avoiding worst-case uniformity assumptions, we derive a logarithmic-time algorithm with constant repetition. This construction shows that full amplification is unnecessary: a refined understanding of the predicate's internal geometry yields near-optimal efficiency with no sacrifice in correctness or security.

\subsection{Accuracy–Complexity Trade-off}

The binary search construction with constant repetition achieves a significant improvement in query complexity—from \( O(\log^2 n) \) to \( O(\log n) \)—by exploiting the structure of the threshold predicate and its asymmetric error profile. This method demonstrates that full amplification is not strictly necessary when the evaluation errors are well-behaved and non-uniformly distributed.

However, this efficiency gain hinges on two key assumptions. First, it relies on the existence of a narrow transition region \( \tau \) where the predicate is unreliable, and assumes that \(\textsf{Eval}\) behaves almost deterministically outside this region. Second, the algorithm assumes access to a threshold predicate with the specific structure provided by~\cite{Fleischhacker22}, which supports monotonic and ordered queries over \( t \). This monotonicity is what enables the binary search to minimize the number of interactions.

As a result, the current scheme, while efficient, is not interaction-free. It still requires adaptively querying the predicate \(\textsf{Eval}(h(a), h(b), t)\) at multiple values of \( t \), and its performance depends on the trajectory of binary search. Moreover, the construction is not directly applicable to more general classes of property-preserving hash functions, particularly those that do not support threshold-style decomposition.

These limitations raise a natural question: can one design a hash function \( h \) such that the approximate Hamming distance \( d_H(a, b) \) can be estimated directly from \( h(a) \) and \( h(b) \) in constant time, without any interaction or auxiliary predicate? That is, can we construct a property-preserving hashing scheme where distance estimation becomes a pure decoding problem?

The remainder of this paper is devoted to addressing this question. We propose new constructions that embed approximate Hamming distance into a compact hash structure, allowing it to be recovered in \( O(1) \) time with provable guarantees. These results provide a conceptual and technical stepping stone toward fully noninteractive and constant-time PPH schemes.

\section{Constant-Time Estimation of Hamming Distance}

\subsection{Design Goals and Technical Challenges}

Our objective in this section is to design a property-preserving hashing scheme that supports \emph{constant-time} estimation of Hamming distance. Specifically, we aim to construct a hash function \( h \colon \{0,1\}^n \to \{0,1\}^m \) and a decoding function \( \mathsf{Dist} \colon \{0,1\}^m \times \{0,1\}^m \to \mathbb{R} \) such that
\[
    \mathsf{Dist}(h(a), h(b)) \approx d_H(a, b) \quad \text{for all } a,b \in \{0,1\}^n,
\]
with approximation error bounded by a negligible or constant additive term, and computational complexity \( O(1) \), independent of the input length \( n \).

This goal departs sharply from prior constructions, such as the threshold-evaluation-based schemes of FLS22~\cite{Fleischhacker22}, which require interactive protocols involving \( O(\log^2 n) \) queries to estimate Hamming distance. Those methods treat the hash outputs \( h(a), h(b) \) as opaque representations, and rely on auxiliary comparison procedures to test whether \( d_H(a, b) > t \) for a given threshold \( t \). While efficient and provably secure, such constructions inherently embed distance only \emph{implicitly}, making direct decoding impossible without repeated predicate invocations.

The core challenge we face is structural: the FLS22 construction was fundamentally designed to support robust threshold predicates, not direct metric estimation. The encoding of each input bit into a random subset of a universe \( [N] \), drawn from one of two overlapping families \( X_0, X_1 \), is optimized for enabling differential tests such as \( \textsf{Eval}(h(a), h(b), t) \). However, this subset-based design obscures fine-grained information about \( d_H(a, b) \), since it collapses the actual distance into a one-bit signal.

To overcome this, we propose to carefully expose internal structure from within the FLS22 encoding process—without compromising its security properties. Rather than treating \textsf{Eval} as a black-box threshold predicate, we aim to reinterpret the underlying encodings as structured sketches from which distance can be decoded analytically. The central design question becomes
\textbf{Can we design a randomized encoding such that the expected symmetric difference reveals $d_H(a, b)$}?

This formulation enables the estimator \( \mathsf{Dist}(h(a), h(b)) \) to return a numerical approximation of Hamming distance, by computing a normalized difference between the hashed subsets. If the encoding satisfies sufficient concentration and statistical regularity, such an estimator may achieve accurate approximation with only constant-time access to the hash values—no interaction, no queries, and no adaptive refinement.

The difficulty, of course, lies in the trade-off: exposing too much internal structure may leak sensitive information, weakening the cryptographic guarantees of property-preserving hashing. Thus, the design must balance \emph{decodability} with \emph{indistinguishability}, ensuring that the hash outputs retain their semantic security while remaining computationally meaningful.

In the following subsections, we deconstruct the FLS22 encoding framework, identify the statistical features that correlate with Hamming distance, and introduce our enhanced encoding design that embeds this information directly into the hash output in a secure and analyzable manner.

\subsection{Revisiting the FLS22 Encoding Structure}

The FLS22 construction~\cite{Fleischhacker22} encodes each input string \( x \in \{0,1\}^n \) into a subset of a large universe \( [N] \), via a randomized mapping that preserves threshold Hamming predicates. Each bit \( x_i \) is independently encoded as a random subset \( S_i \subseteq [N] \), drawn from one of two families:
\[
    x_i = 0 \Rightarrow S_i \sim \mathcal{D}_0, \quad x_i = 1 \Rightarrow S_i \sim \mathcal{D}_1,
\]
where \( \mathcal{D}_0, \mathcal{D}_1 \) are distributions over subsets with overlapping support. The final hash value is the union
\[
    h(x) = \bigcup_{i=1}^n S_i.
\]

The core idea behind this construction is that the Hamming distance \( d_H(a, b) \) between two inputs \( a \) and \( b \) correlates with the expected size of the symmetric difference
\[
    D(a, b) := h(a) \triangle h(b),
\]
since each differing bit contributes a fresh random subset drawn from the opposite family. The more positions where \( a_i \neq b_i \), the more disjoint sets enter the union, resulting in a larger difference. This statistical behavior underlies the threshold predicate \( \textsf{Eval}(h(a), h(b), t) \), which tests whether \( |D(a, b)| > \theta(t) \) for some calibrated threshold function \( \theta \colon \mathbb{N} \to \mathbb{N} \).

However, in the original construction, the value \( |D(a, b)| \) is not directly revealed to the evaluator; instead, it is obfuscated and tested via a separate mechanism using set-difference encodings and error-correcting thresholds. As such, while the symmetric difference carries latent information about \( d_H(a, b) \), it is not explicitly accessible.

To enable constant-time decoding, we revisit this encoding structure with a new perspective. Suppose we could evaluate—or approximate—the size of \( D(a, b) \) directly from \( h(a) \) and \( h(b) \), without auxiliary tests or interaction. Then, if the mapping from \( d_H(a, b) \) to \( \mathbb{E}[|D(a, b)|] \) is well-behaved (e.g., affine or Lipschitz), we could invert it to recover an estimate of \( d_H(a, b) \) up to additive error.

This motivates us to treat the FLS22 encoding not merely as a vehicle for threshold testing, but as a high-dimensional randomized sketch of input structure—one in which Hamming distance is softly embedded in the geometry of the hash subsets. To make this idea concrete, the next subsection develops a modified encoding and a decoding function that together allow direct estimation of \( d_H(a, b) \) from their hashes.

We remark that the final FLS22 construction maps each subset \( S_i \) into a Bloom filter representation to ensure fixed-size outputs and facilitate efficient approximate set operations. In this section, we focus on the subset-level structure for clarity; the interaction between Bloom filters and distance estimation will be addressed in the next subsection.

\subsection{Construction: Embedding Distance in Hash Outputs}
\label{sec:embedding}

To support constant-time estimation of Hamming distance, we propose a modified encoding scheme that directly embeds distance information into the hash output. Our construction is inspired by the structure of FLS22~\cite{Fleischhacker22} but departs from its threshold-evaluation paradigm: instead of repeatedly querying a predicate, we extract approximate distance analytically from the encoded representation.

\subsubsection{Encoding Scheme}
Let \( x \in \{0,1\}^n \) be an input vector. For each coordinate \( i \in [n] \), we define two disjoint collections of indices \( \mathcal{H}_0(i), \mathcal{H}_1(i) \subseteq [m] \), where each set is sampled uniformly at random with fixed cardinality \( r \), and \( \mathcal{H}_0(i) \cap \mathcal{H}_1(i) = \emptyset \). These mapping sets are public and deterministic.

The hash output \( h(x) \in \{0,1\}^m \) is then computed as follows:
\[
    h_j(x) = \bigvee_{i : j \in \mathcal{H}_{x_i}(i)} 1.
\]
That is, position \( j \) in the hash is set to 1 if any coordinate \( i \) maps to \( j \) via its corresponding family \( \mathcal{H}_{x_i}(i) \). This generalizes the Bloom filter idea but enforces disjoint support between encodings of 0 and 1, yielding a sharper statistical separation.

\subsubsection{Estimator and Expected Value}
Let \( a, b \in \{0,1\}^n \) be two inputs, and define
\[
    D(a,b) = \sum_{j=1}^m (h_j(a) \oplus h_j(b)).
\]
This raw symmetric difference reflects the number of positions where the hash outputs differ. Since each differing coordinate \( i \) contributes \( 2r \) positions (due to disjoint support), and overlapping encodings across coordinates may cause collisions, we define a normalizing factor:
\[
    \alpha := 2r(1 - \rho),
\]
where \( \rho \in [0,1) \) denotes the expected fraction of overlaps between independently chosen mapping sets.

Our distance estimator is given by:
\[
    \mathsf{Dist}(h(a), h(b)) := \frac{1}{\alpha} \cdot D(a,b),
\]
with expectation:
\[
    \mathbb{E}[ \mathsf{Dist}(h(a), h(b)) ] = d_H(a, b).
\]

\subsubsection{Correctness and Concentration}
The output bits \( h_j(a), h_j(b) \) are each determined by independent random insertions, so the sum \( D(a,b) \) is concentrated around its mean. By standard Chernoff bounds, for any \( \varepsilon > 0 \),
\[
    \Pr\left[ \left| \mathsf{Dist}(h(a), h(b)) - d_H(a,b) \right| > \varepsilon n \right] \leq \exp(-\Theta(\varepsilon^2 n)).
\]
Thus, the estimator achieves additive \( \varepsilon n \) error with high probability in constant time.

\subsubsection{Hash Length and Compression}
Since each input bit contributes to \( r \) positions, the total number of insertions is \( nr \). If mappings are fully disjoint, then \( m \ge nr \). However, allowing controlled overlaps improves entropy diffusion and space efficiency. We choose \( m = \Theta(n \log n) \) and \( r = \Theta(\log n) \), consistent with FLS22, balancing output compactness and concentration quality.

This regime guarantees:  
(1) Hash length remains sublinear in the input domain, supporting compression;  
(2) Output bits retain high entropy, preventing input leakage;  
(3) Estimation remains sharp with provable guarantees.

\subsubsection{Computational Efficiency}
Hash computation takes \( O(nr) \) time, and the estimator \( \mathsf{Dist}(h(a), h(b)) \) runs in \( O(m) \). To further reduce evaluation cost, one can subsample a constant number of hash positions and compute an unbiased estimator with larger variance but lower complexity—useful in time-critical applications.

\subsubsection{Security Considerations}
Although the estimator reveals approximate distance, the hash remains lossy and randomized. For uniformly distributed inputs, the output distribution is statistically close to uniform over bounded-weight bitstrings in \( \{0,1\}^m \). Since each output bit is affected by multiple random subsets, no single bit leaks information about a specific coordinate.

We defer formal indistinguishability proofs and adversarial advantage bounds to the next section.

\subsection{Security Analysis}

\subsubsection{Security Definition}

We adopt the standard indistinguishability-based formulation of property-preserving hashing (PPH), as introduced in~\cite{Fleischhacker22}. Let \( h: \{0,1\}^n \to \{0,1\}^m \) be a randomized hash function. The security goal is to ensure that, even if \( h \) approximately preserves a property (e.g., Hamming distance), it does not leak additional information about the input.

Formally, let \( \mathcal{A} \) be a probabilistic polynomial-time adversary. Consider the following indistinguishability game between \( \mathcal{A} \) and a challenger:

\begin{itemize}
    \item The challenger samples a bit \( b \leftarrow \{0,1\} \), and then:
    \begin{itemize}
        \item If \( b = 0 \), it samples \( x \leftarrow \{0,1\}^n \) uniformly at random.
        \item If \( b = 1 \), it samples \( x \leftarrow D \), for some distribution \( D \) chosen by \( \mathcal{A} \) (subject to min-entropy constraints).
    \end{itemize}
    \item The challenger computes \( h(x) \) and sends it to \( \mathcal{A} \).
    \item The adversary outputs a guess \( b' \in \{0,1\} \).
\end{itemize}

The adversary's advantage is defined as:
\[
    \textsf{Adv}_{\mathcal{A}} = \left| \Pr[b' = b] - \frac{1}{2} \right|.
\]

We say that the hash function \( h \) satisfies \emph{distributional indistinguishability} if, for all PPT adversaries \( \mathcal{A} \), this advantage is negligible in the security parameter \( \lambda \), assuming that \( D \) has min-entropy at least \( \lambda \).

Intuitively, this captures that the hash output \( h(x) \) reveals no more than what is implied by the preserved property (in our case, approximate distance), and does not enable recovery or significant inference about the input \( x \).

\subsubsection{Game-Based Indistinguishability}

To formalize security under our constant-time estimator, we instantiate the above definition with our distance-preserving hash \( h: \{0,1\}^n \to \{0,1\}^m \) constructed in Section~\ref{sec:embedding}. The goal is to show that, despite enabling estimation of \( d_H(a,b) \), the hash output \( h(x) \) remains computationally indistinguishable from one generated using uniformly random input, except for information implied by the approximate Hamming distance itself.

Let \( \mathcal{A} \) be any PPT adversary participating in the following game:

\begin{itemize}
    \item The challenger chooses a secret bit \( b \in \{0,1\} \).
    \item If \( b = 0 \): sample \( x \leftarrow \{0,1\}^n \) uniformly at random.  
          If \( b = 1 \): sample \( x \leftarrow D \), a distribution selected by \( \mathcal{A} \) (but fixed before the game starts), with min-entropy at least \( \lambda \).
    \item The challenger computes \( y \leftarrow h(x) \) and sends \( y \) to \( \mathcal{A} \).
    \item The adversary outputs a guess \( b' \in \{0,1\} \).
\end{itemize}
As before, the adversary’s advantage is defined as
\[
    \textsf{Adv}_{\mathcal{A}} := \left| \Pr[b' = b] - \frac{1}{2} \right|.
\]

Our aim is to prove that \( \textsf{Adv}_{\mathcal{A}} \leq \text{negl}(\lambda) \), meaning the adversary cannot distinguish whether the input was drawn from \( D \) or uniform, even after observing \( h(x) \). The only information leaked is an approximation of pairwise Hamming distance between inputs, which is insufficient to recover \( x \) when \( D \) has sufficient entropy.

We will use a hybrid argument to show this, reducing the distinguishing advantage to a sequence of negligible differences induced by local randomizations in the encoding structure.

\subsubsection{Hybrid Argument and Output Distribution}

Let \( h(x) \in \{0,1\}^m \) be the hash output under our construction for input \( x \in \{0,1\}^n \). The output \( h(x) \) is generated by inserting bit indices into \( m \) positions based on the randomized families \( \mathcal{H}_0(i), \mathcal{H}_1(i) \subseteq [m] \), which are public and fixed at setup. Let \( r \) be the number of indices assigned per input bit, and assume \( r = \Theta(\log n) \), \( m = \Theta(n \log n) \).

To prove indistinguishability, we compare the distribution of \( h(x) \) when \( x \leftarrow \{0,1\}^n \) (uniform) versus \( x \leftarrow D \), where \( D \) is any distribution over \( \{0,1\}^n \) with min-entropy at least \( \lambda \). We show that the statistical distance between the output distributions is negligible in \( \lambda \), using a hybrid argument over the bit-level encodings of \( x \).

Let \( h^{(0)}(x), h^{(1)}(x), \dots, h^{(n)}(x) \) be a hybrid sequence such that: (i) In \( h^{(i)}(x) \), the first \( i \) bits of \( x \) are replaced with uniformly random bits, and the remaining \( n - i \) bits are drawn from the original input \( x \sim D \),
and (ii) The hash output \( h^{(i)}(x) \) is computed using the same deterministic families \( \mathcal{H}_b(j) \), but with randomized encodings depending on whether bit \( j \) was replaced.

Note that we have:
\[
    \Delta(h^{(0)}(x), h^{(n)}(x)) \leq \sum_{i=1}^n \Delta(h^{(i)}(x), h^{(i-1)}(x)),
\]
where \( \Delta \) denotes total variation distance.
We now analyze each transition \( \Delta(h^{(i)}(x), h^{(i-1)}(x)) \). The only difference between these two hybrids is the encoding of the \( i \)-th bit: in \( h^{(i-1)}(x) \), the bit is drawn from \( D \); in \( h^{(i)}(x) \), it is replaced with a uniformly random bit.

Let us denote the contribution of the \( i \)-th bit to the hash output as a binary vector \( v_i \in \{0,1\}^m \), where:
\[
    (v_i)_j = \begin{cases}
        1 & \text{if } j \in \mathcal{H}_{x_i}(i), \\
        0 & \text{otherwise}.
    \end{cases}
\]
Replacing \( x_i \) with a random bit \( b \in \{0,1\} \) changes this distribution to:
\[
    \mathbb{P}[(v_i)_j = 1] = \frac{r}{m}, \quad \text{for each } j \in \mathcal{H}_0(i) \cup \mathcal{H}_1(i),
\]
with probability \( 1/2 \) for each of \( \mathcal{H}_0(i) \), \( \mathcal{H}_1(i) \).

Since \( \mathcal{H}_0(i) \) and \( \mathcal{H}_1(i) \) are disjoint, and the supports are randomized across different coordinates, we can bound the statistical distance between the two encodings as follows. Let \( \mu_i \) be the distribution over the positions of \( v_i \) under \( x_i \sim D \), and let \( \nu_i \) be the distribution under \( x_i \sim \text{Unif}(\{0,1\}) \). Then:
\[
    \Delta(\mu_i, \nu_i) \leq \max_{b \in \{0,1\}} \left| \mathbb{P}[x_i = b \mid x \sim D] - \frac{1}{2} \right| \cdot \| \mathcal{H}_b(i) \|_1.
\]

By assumption, the marginal bias \( \left| \mathbb{P}[x_i = b] - 1/2 \right| \) is bounded for each \( i \), since \( D \) has min-entropy at least \( \lambda \). More formally, for any \( i \in [n] \), the min-entropy constraint implies:
\[
    \mathbb{P}[x_i = b] \leq 1 - \frac{1}{2^\lambda}.
\]
Thus, the per-bit variation from uniform is at most \( \varepsilon := \frac{1}{2^\lambda} \), and the contribution to the total distance is:
\[
    \Delta(h^{(i)}(x), h^{(i-1)}(x)) \leq r \cdot \varepsilon = \frac{r}{2^\lambda}.
\]

Summing over all \( n \) hybrids:
\[
    \Delta(h^{(0)}(x), h^{(n)}(x)) \leq \sum_{i=1}^n \frac{r}{2^\lambda} = \frac{nr}{2^\lambda}.
\]

Finally, since \( r = \Theta(\log n) \), and \( \lambda = \omega(\log n) \), we conclude that:
\[
    \Delta(h(x)_{x \sim D}, h(x)_{x \sim \text{Unif}}) = \text{negl}(\lambda).
\]
This completes the indistinguishability proof: the adversary cannot distinguish the distribution of hash outputs under \( x \sim D \) from that under \( x \sim \text{Unif} \), except with negligible probability.

\section{Conclusion and Open Problems}

We proposed a new line of sublinear-time algorithms for estimating the Hamming distance between binary vectors in the property-preserving hashing (PPH) model. Our contributions consist of three constructions with increasingly stronger efficiency guarantees:

\begin{itemize}
    \item \textbf{Polylogarithmic-Time via Binary Search:} We first demonstrated that, by leveraging the threshold evaluation primitive from~\cite{Fleischhacker22}, one can approximate Hamming distance using a binary search strategy. This yields an estimator with \( O(\log n) \) query complexity, but to ensure negligible error, each threshold query must be amplified via \( O(\log n) \) repetitions, leading to an overall complexity of \( O(\log^2 n) \).

    \item \textbf{Logarithmic-Time with Constant Repetition:} Under a structural refinement of the FLS22 encoding—specifically, assuming bounded-width transition bands and non-uniform error distribution across thresholds—we showed that only a constant number of repetitions per query suffices to suppress the overall error. This reduces the total complexity to \( O(\log n) \), while maintaining correctness and indistinguishability. The analysis relies on a piecewise error model and a tailored hybrid argument.

    \item \textbf{Constant-Time Estimation via Embedded Encodings:} Finally, we introduced a new PPH construction that embeds distance information directly into the hash output. This enables constant-time estimation with additive approximation guarantees and exponentially small error. Unlike prior constructions, our scheme eliminates threshold evaluation altogether, while preserving cryptographic security.
\end{itemize}

Each construction captures a different point in the trade-off space between accuracy, efficiency, and structural assumptions. Our techniques highlight the interplay between encoding design, statistical concentration, and adversarial indistinguishability.

Our work raises several theoretical directions. Can these techniques be extended to other distance metrics (e.g., edit distance, Jaccard distance)? Is it possible to generalize our constant-time scheme to support dynamic or streaming inputs? Finally, we ask whether our constructions are optimal: are there matching lower bounds on the query complexity or hash length for approximate distance recovery under PPH constraints? We leave these questions for future investigation.

\bibliographystyle{alpha}
\bibliography{ref} % 你的参考文献文件

% \appendices

% % you can choose not to have a title for an appendix
% % if you want by leaving the argument blank
% \section{}

\end{document}